\documentclass{ifac/ifacconf}

\usepackage{graphicx,xcolor}      % include this line if your document contains figures
\usepackage{natbib}        % required for bibliography
\usepackage{comment}
\usepackage{amsfonts}
\usepackage{amsmath}
\usepackage{mathtools}
\usepackage{subcaption}
\usepackage{lipsum}
\usepackage{tikz}
\usepackage{float}
\usepackage{url}
\usepackage{booktabs}

\usepackage{caption}

\definecolor{color1}{HTML}{2E15C8}
\definecolor{color2}{HTML}{BE37A4}
\definecolor{color3}{HTML}{F58275}
\definecolor{color4}{HTML}{FFD700}

\DeclareMathOperator*{\argmax}{arg\,max}    % argmax
\begin{document}
\begin{frontmatter}

%\title{Communication-Efficient Control\\ of Multi-Agent Formations\thanksref{footnoteinfo}} 
\title{VoI-aware Scheduling Schemes for Multi-Agent Formation Control\thanksref{footnoteinfo}}

% Title, preferably not more than 10 words.

\thanks[footnoteinfo]{The authors contributed equally to this work, and they are listed in alphabetical order. This work was partially carried out within the Italian National Center for Sustainable Mobility (MOST) and received funding from NextGenerationEU (Italian NRRP – CN00000023 - D.D. 1033 17/06/2022 - CUP C93C22002750006).}

\author{Federico Chiariotti,} 
\author{Marco Fabris} 

\address{Department of Information Engineering, University of Padova, 
   Italy\\(e-mail: federico.chiariotti@unipd.it, marco.fabris.1@unipd.it).}
%\address[Second]{Colorado State University, 
%   Fort Collins, CO 80523 USA (e-mail: author@lamar. colostate.edu)}
%\address[Third]{Electrical Engineering Department, 
%   Seoul National University, Seoul, Korea, (e-mail: author@snu.ac.kr)}

% PIN Federico: 179151
% PIN Marco: 126608
\begin{abstract}                % Abstract of 50--100 words
Formation control allows agents to maintain geometric patterns using local information, but most existing methods assume ideal communication. This paper introduces a goal-oriented framework combining control, cooperative positioning, and communication scheduling for first-order formation tracking. Each agent estimates its position using 6G network-based triangulation, and the scheduling of information updates is governed by Age of Information (AoI) and Value of Information (VoI) metrics. We design three lightweight, signaling-free scheduling policies and assess their impact on formation quality. Simulation results demonstrate the effectiveness of the proposed approach in maintaining accurate formations with no additional communication overhead, showing that worst-case formation adherence increases by $20\%$.

\end{abstract}

\begin{keyword}
Value of Information, Age of Information, Multi-Agent Systems, Formation Control
\end{keyword}

\end{frontmatter}
%===============================================================================

\section{Introduction}\label{sec:intro}

Formation control concerns the coordination of multiple autonomous agents to achieve and maintain a desired geometric configuration using local sensing and communication. This capability has broad applications ranging from Unmanned Aerial Vehicle (UAV) swarms~(\cite{anderson2008uav,wang2024distributed}) to mobile and visual sensor networks~(\cite{wang2005formation,varotto2022visual}) to space missions and self-organizing robotic teams~(\cite{beard2001coordination,Fabris2019coverage}.) A key requirement in formation control is to ensure the system's scalability, robustness, and autonomy under %distributed
limiting conditions where each agent has access only to local measurements.

Early formation control relied on global information or centralized methods, which lack scalability and are prone to single-point failures. Recent distributed schemes rely on local interactions over sensing or communication graphs, inspired by biological swarms~(\cite{reynolds1987flocks}.) This shift was driven by consensus approaches~(\cite{olfati2007consensus},) which guide system dynamics toward a desired shape, and rigidity theory~(\cite{thorpe1999rigidity},) which ensures formation uniqueness and local stability~(\cite{de2019formation}.)

%Extensions to directed formations, such as persistent graphs, allow for convergence in cases with asymmetrical information exchange \cite{hendrickx2008formation}.

%Gradient-based formation control is among the most widely studied decentralized strategies. It leverages potential functions based on inter-agent distances, allowing agents to converge to the desired formation using only local relative information. Rigidity theory provides the guarantees needed to ensure the uniqueness and local stability of such formations \cite{zelazo2019formation,de2019formation}. Extensions to directed formations, such as persistent graphs, allow for convergence in cases with asymmetrical information exchange \cite{hendrickx2008formation}.

Recent advances include area and bearing constraints to enhance flexibility and observability~(\cite{trinh2018bearing,zhao2019bearing,Fabris2022bearing},) and the integration of orientation alignment to compensate for the lack of a global frame. Embedding orientation estimation~(\cite{bertoni2022indoor},) distributed algorithms for the estimation of relative quantities~(\cite{garin2010survey,Fabris2022SERM}) or secure consensus protocols~(\cite{Fabris2022robustconsensus,Fabris2023structuredconsensus}) enables agents to safely align local frames and achieve improved global convergence. Formation tracking, where agents preserve a geometric shape while following a trajectory, has also emerged as a key extension~(\cite{ Fabris2019formation1,FABRIS2024OIFT},) often relying on distributed online estimators to compute the formation centroid~(\cite{antonelli2012decentralized}.)
% Fabris2019formation1,

Nonetheless, most formation control frameworks assume ideal communication, yet real systems face issues like packet loss, bandwidth limits, and delays. Although relatively limited, some works address these challenges: see, for instance,  \cite{cepeda2016stability,hoff2024communication}. A growing research focus is on incorporating communication constraints into control design. Hybrid strategies combining estimation and control have shown effectiveness under intermittent communication--see, e.g., \cite{bartels2016robust,zhao2020vehicle}--while robust methods for partial information remain an open challenge.

The impossibility to disentangle control and communication in all but the simplest systems has been known since the famous counterexample proposed by~\cite{witsenhausen1968counterexample}. The development of networked position systems (see~\cite{yang2024positioning}) has reached centimeter-level accuracy, and the integration of 5G and 6G networks' sensing capabilities with UAV and robotic applications is a major technology driver, as outlined by~\cite{fan2024air}.

On the other hand, communication systems have always been designed based on fixed requirements and simplistic application models, and goal-oriented communications, which consider the performance of the task the communication is meant to help, as described by~\cite{gunduz2022beyond}, represent a very recent development. The introduction of Age of Information (AoI) (see~\cite{yates2021age}) which is more relevant to control applications than mere network latency, about a decade ago has paved the way for more integrated approaches. 

The more recent development of scheduling approaches based on Value of Information (VoI), as described in detail by~\cite{alawad2022value} has further reduced the gap between communication metrics and control requirements, as VoI can represent arbitrary loss functions. Indeed, several recent VoI-aware scheduling schemes, see, e.g.,~\cite{holm2023goal,wang2024value}, show significant performance gains, and goal-oriented approaches are a major technology driver towards the development of 6G, as described by~\cite{gunduz2023timely}.

Furthermore, a recent work by~\cite{talli2025pragmatic} has shown that reaching the optimal joint solution is computationally infeasible in the general case of Markov Decision Processes (MDPs) with distributed observations and limited communication resources. However, the optimal solution is reachable if the control problem has a particular structure, as shown by~\cite{zheng2020urgency} for linear problems.
The gap between the goal-oriented communication literature and the vast body of work on distributed control with local observations is still significant, and the integration of existing distributed approaches with VoI-based scheduling and resource allocation is still mostly unexplored.

In this work, we apply goal-oriented communication concepts to a formation control task: we consider a multi-agent control algorithm, in which autonomous agents need to maintain a formation while following a predetermined trajectory, and optimize network-based positioning to improve its accuracy. In the proposed scheme, individual agents use the 6G network to triangulate their position, and AoI- and VoI-based scheduling is used to guarantee the adherence to the formation while limiting the amount of communication resources used by the formation. 

The main contributions of this paper are the following:
\begin{enumerate}
    \item We propose a joint control and cooperative positioning framework for multi-agent formation control in $3$ dimensions;
    \item We design three scheduling policies based on AoI and VoI that can be pre-computed and implemented without additional signaling;
    \item We analyze the impact of the scheduling policies on the quality of the formation, showing that the formation loss function's $99$th percentile decreases by $20\%$ when using a VoI-aware scheduling policy.
\end{enumerate}

The rest of the paper is organized as follows: first, Sec.~\ref{sec:prelim} presents the system model and theoretical background. The scheduling policies and control method are then described in detail in Sec.~\ref{sec:method}, and Sec.~\ref{sec:exp} presents our simulation results. Finally, Sec.~\ref{sec:conc} concludes the paper and presents some possible avenues for future work.

\section{Theoretical background and models}\label{sec:prelim}

In this section, the models and assumptions are provided to formalize the problem. In addition, communication topology constraints are taken into account for the development of a related distributed control law.

\subsection{Notation and multi-agent network model}

 We introduce the notation used throughout the paper.
The sets of natural and real numbers are respectively indicated by $\mathbb{N}$ and $\mathbb{R}$. 
The symbols $|\cdot|$ and $\left\|\cdot \right\|$ denote the set cardinality and Euclidean norm, respectively. % and zero matrix having dimensions that will be clear from the context. 
%\mfa{@Marco: SE SERVE we define the vectorization operator and the diagonal operator as ... }
Lastly, the variable $t\geq 0$ specifies the continuous time instants, while $\mathbb{E}$ denotes the expected value. %a trajectory as $t$ spans the interval $[0,T]$, for a chosen $T>0$. %\mfa{@Marco: SE SERVE Symbols $\nabla_{*}$ and $\mathcal{H}_{**}$ indicate standard gradient and Hessian operators w.r.t. some generic variable~$*$. }

In this work we consider multi-agent systems that can be topologically represented by an undirected graph $\mathcal{G} = (\mathcal{V},\mathcal{E},\mathcal{W})$, consisting of the set of nodes $\mathcal{V}$, the set of edges $\mathcal{E}$ and the set of weights $\mathcal{W}$ (see also \cite{mesbahi2010graph} for a detailed discussion on multi-agent network modeling). Each node of $\mathcal{V}$ is simply addressed by its index $i = 1,\ldots,n$, where $n = |\mathcal{V}|$ is the cardinality of $\mathcal{V}$, and it is weighted by $a_{ii} \in \mathcal{W}$, with $a_{ii}> 0$. The node $i$ is referred to as the $i$-th agent. Also, node $i$ is allowed to exchange information with node $j$ if and only if there exists an edge $(i,j)=(j,i)$ contained in $\mathcal{E}$. Each edge $(i,j)\in \mathcal{E}$ is weighted by $a_{ij} \in \mathcal{W}$, with $a_{ij}>0$.
We assume that $\mathcal{G}$ is connected, i.e. there exists a collection of edges that link each node $i$ to any other node $j \neq i$. The neighborhood of node $i$ is defined as the set $\mathcal{N}_{i} = \{j\in \mathcal{V}|(i,j)\in \mathcal{E}\}$. 
%The %Laplacian matrix associated to $\mathcal{G}$ is defined as $\mathbf{L}=\mathbf{D}-\mathbf{A}$, where $\mathbf{A}\in \mathbb{R}^{n\times n}$ denotes the 
%adjacency matrix of $\mathcal{G}$ is such that $[A]_{ij} = 1$ if $(i,j) \in \mathcal{E}$; $[A]_{ij} = 0$ otherwise, whereas $D\in \mathbb{R}^{n\times n}$ denotes the degree matrix of $\mathcal{G}$, such that $D$ is diagonal and $[D]_{ii} = \sum_{j=1}^{n} [A]_{ij}$.

%\subsection{Agents' dynamics}
We suppose that $n\geq 2$ robotic agents with linear dynamics are already deployed in an environment space of dimension $d$, such that $d\in \left\lbrace 1,2,3\right\rbrace$. 
Each agent $i$ also maintains an estimate of its absolute position $p_{i}(t) \in \mathbb{R}^{d}$, which we denote as $\hat{p}_{i}(t) \in \mathbb{R}^{d}$. The agent can control its position by means of a regulation on its absolute velocity $v_{i}(t)\in \mathbb{R}^{d}$. We assume that control is imperfect: the real velocity of the node is
\begin{equation}\label{eq:imperfect_vel}
    \dot{p}_i(t)=v_i(t)+w_i(t),
\end{equation}
where $w_i(t)\sim\mathcal{N}(0,\sigma_i^2 I_d)$ is an additive Gaussian noise.
In the next lines, we drop the dependence of position and velocity on the time $t$ whenever the context allows it.
Setting $N=nd$, the expressions of the estimated state $\hat{p}(t)\in \mathbb{R}^{N}$ and the input $v(t)\in \mathbb{R}^{N}$ for this group of mobile elements are provided respectively by
\begin{equation}
	\hat{p} = \begin{bmatrix}
		\hat{p}_{1}^{\top} & \dots & \hat{p}_{n}^{\top} 
	\end{bmatrix}^{\top},\quad
	v = 
	\begin{bmatrix}
        v_{1}^{\top} & \dots & v_{n}^{\top}
	\end{bmatrix}^{\top}.
\end{equation}

\subsection{Positioning and communication model}

We assume that there are two communication networks: the first is an infrastructure-based 6G connections to multiple Base Stations (BSs) placed in the area, while the second is a short-range device-to-device communication technology operating over a different band, as described by~\cite{azari2020uav}.

The information on the system centroid  $\hat{p}_c(t) \in \mathbb{R}^d$, where
\begin{equation}
    \hat{p}_c = n^{-1} \sum\nolimits_{i=1}^n \hat{p}_i,
\end{equation}
is estimated by the agents through consensus (see also \cite{bullo2018lectures} for a detailed discussion on the topic) and it is thus made locally accessible for each agent running over the short-range connections between the agents. For this reason, the communication network graph $\mathcal{G}$ among the agents is assumed to be connected. 

\begin{figure}[t]
    \centering   
    \includegraphics[width=0.3\textwidth]{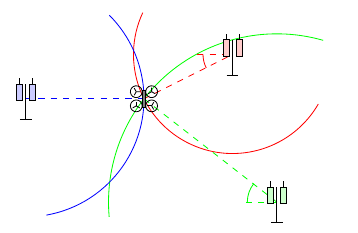}
    \caption{A schematic of a time and angle of arrival based network-based triangulation process.}
    \label{fig:triangulation}
\end{figure}

As the velocity of the agents in~\eqref{eq:imperfect_vel} is imperfect, the system will tend to accumulate noise over time: the difference between a given coordinate of $p_i$ and $\hat{p}_i$ follows a Wiener process, scaled by $\sigma_i$. Consequently, the error over the position after a period $T$ without any updates follows a $\chi^2$ distribution with $d$ degrees of freedom, scaled by $T\sigma_i^2$:
\begin{equation}\label{eq:error}
    \|p_i-\hat{p}_i\|^2T^{-1}\sigma_i^{-2}\sim\chi^2(d),
\end{equation}
where $\|x\|$ represents the Euclidean norm of vector $x$.

We then consider a setup in which a node can periodically broadcast a wireless signal to a number of fixed BSs in the area, as shown in Fig.~\ref{fig:triangulation}. The BSs can then use different techniques relying on the time or angle of arrival of the signal, or on the received power, to triangulate the agent's position. For an overview of these techniques and their evolution, we refer the reader to~\cite{yang2024positioning}. One of the BSs can then broadcast the position to the agents.

As multiple techniques with centimeter-level accuracy are available, we assume that broadcasting the position resets the estimated position $\hat{p}_i(t)$ to the real value $p_i(t)$. However, this operation requires the use of precious communication resources, as wideband signals result in a higher localization accuracy, as discussed by~\cite{behravan2022positioning}.

We then consider a system in which a single node may ask for localization every $T_s$ seconds, to reduce the burden on the network, which might also be required to transmit data from the agents' cameras and sensors, as well as the regular data traffic over the area. The identity of the sensor whose position is to be measured can be determined through a scheduling mechanism, whose definition is at the core of the proposed technique in this paper.

\section{Proposed method}\label{sec:method}

In this section, we propose a method that combines the scheduling of localization services based on a 6G network under bandwidth resource limitations with distributed first-order formation tracking control.

\subsection{Distributed formation tracking control law}

Let us define the desired interagent distances $d_{ij} > 0$, for which it is required that the underlying formation satisfy
\begin{equation}\label{eq:def_formationgoal}
    \lim_{t\rightarrow +\infty} \| \hat{p}_i(t)-\hat{p}_j(t) \| = d_{ij}, \quad \forall (i,j) \in \mathcal{V} \times \mathcal{V}.  
\end{equation}
We assume that the bar-and-joint framework $\mathcal{F}(\mathcal{G}):=(\hat{p},\{d_{ij}\}_{(i,j)\in \mathcal{E}},\mathcal{G})$ is infinitesimally rigid (\cite{de2019formation},) so that only the subset comprising the desired distances $d_{ij}$ such that $(i,j) \in \mathcal{E}$ can be considered while attaining the configuration prescribed by~\eqref{eq:def_formationgoal}.
Furthermore, let us denote with $p_{c,\text{des}}(t)\in \mathbb{R}^d$ the desired trajectory to be tracked by the formation centroid $p_c(t)$. Since the latter quantity is global, it is worth to observe that a strategy to estimate the formation centroid is needed. In this direction, each agent is thus endowed with an additional local state $q^i(t) \in \mathbb{R}^N$, where $q^i = \begin{bmatrix}
    (q_1^i)^\top  & \cdots & (q_n^i)^\top
\end{bmatrix}^\top$, $q_j^i \in \mathbb{R}^d$. Such a state augmentation allows to estimate the global position information $\hat{p}$ through, e.g., the consensus protocol devised by \cite{antonelli2012decentralized} 
\begin{align}\label{eq:Antonelliest}
        \dot{q}^i &= -K_E \sum_{j\in \mathcal{N}_i} a_{ij} (q^i-q^j) - K_E a_{ii} N_i(q^i - \hat{p}) + v^i ,
\end{align}
where $K_E > 0$ is a tunable estimation gain and $N_i$ is a selection matrix for which only the components of $q^i_i$ are chosen in $q^i$, while the remaining ones are zeroed. Also, in \eqref{eq:Antonelliest}, the input
 $v^i(t)\in \mathbb{R}^N$, $v^i = \begin{bmatrix}
    (v^i_1)^\top & \cdots & (v^i_n)^\top
\end{bmatrix}^\top$, $v_j^i \in \mathbb{R}^d$, is set as follows:
\begin{align}
    v^i_j = \mathcal{I}\left(j\in \mathcal{N}_i \cup \{i\}\right)v_j,
\end{align}
where $\mathcal{I}(\cdot)$ is the indicator function, which is equal to $1$ if the condition inside the argument is true and $0$ otherwise. The value of~\eqref{eq:Antonelliest} is initialized as in~\cite{FABRIS2024OIFT}:
\begin{align}
    q^i_j(0) = \mathcal{I}\left(j \notin \mathcal{N}_i \cup \{i\}\right)
        (1+|\mathcal{N}_{i}|)^{-1} \sum_{\mathclap{k \in \mathcal{N}_i \cup \{i\}}} \hat{p}_k(0).
\end{align}

Exploiting this distributed approach, each agent in the network can efficiently compute a local estimate $\hat{p}^i_c(t)\in \mathbb{R}^d$ of the formation centroid $\hat{p}_c(t)$ at any instant $t$ as
\begin{equation}
    \hat{p}^i_c = n^{-1} \sum\nolimits_{j=1}^n q^i_j.
\end{equation}
In light of the above premise, to steer the formation under consideration we choose the following well-known distributed stabilizing gradient-based law (\cite{ahn2020formation}):
\begin{align}
    v_i =~ & -K_P a_{ii}(\hat{p}_c^i-p_{c,\text{des}}) \nonumber\\
    & -K_F  \sum_{j\in \mathcal{N}_i}a_{ij} (\|\hat{p}_i-\hat{p}_j\|^2 - d_{ij}^2)(\hat{p}_i-\hat{p}_j),
\end{align}
where $K_P>0$ and $K_F>0$ are control gains that can be tuned\footnote{To perform tracking properly, $K_E \gg K_F,K_P$ should be selected.} to effectively accomplish the tracking and formation tasks, respectively.

\begin{table*}[h]
	\captionsetup{font=small, width=\textwidth}
	\caption{Edge sets of the considered formations.}
	\label{tab:edgesets}
    \scriptsize 
    \centering
	\begin{tabular}{lc}
		\toprule
		Formation type & Edge set $\mathcal{E}$ \\
		\midrule
        Symmetrical & \{(1,2),(1,4),(1,5),(1,6),(1,8),(2,3),(2,4),(2,6),(2,8),(3,4),(3,5),(3,6),(3,7),(4,6),(4,8),(5,6),(5,7),(5,8),(6,7),(7,8)\}\\
        Asymmetrical & \{(1,2),(1,4),(1,5),(1,6),(1,8),(2,3),(2,4),(2,6),(2,7),(2,8),(3,4),(3,5),(3,7),(3,8),(4,7),(4,8),(5,6),(5,7),(5,8),(6,7),(6,8),(7,8)\}\\
		\bottomrule
	\end{tabular}
\end{table*}

\begin{figure*}[t]
    \centering
    \begin{subfigure}[b]{0.3\textwidth}  % Adjust width as needed
        \centering
        \includegraphics[width=\textwidth]{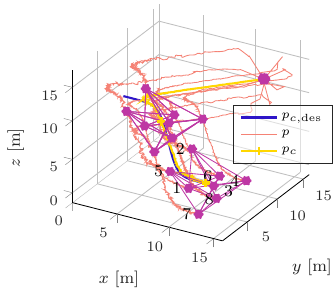}
        \caption{Oracle scheduler.}
        \label{fig:traj_asym_or}
    \end{subfigure}
    \begin{subfigure}[b]{0.3\textwidth}  % Adjust width as needed
        \centering
        \includegraphics[width=\textwidth]{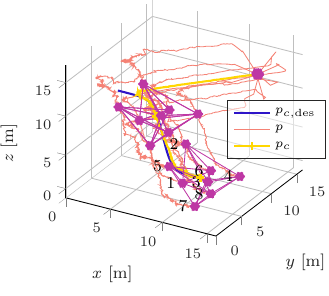}
        \caption{MAF scheduler.}
        \label{fig:traj_asym_maf}
    \end{subfigure}
    \begin{subfigure}[b]{0.3\textwidth}  % Adjust width as needed
        \centering
        \includegraphics[width=\textwidth]{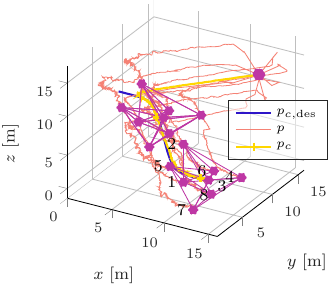}
        \caption{MV scheduler.}
        \label{fig:traj_asym_mv}
    \end{subfigure}
    \caption{3D asymmetrical formation trajectories with different schedulers. In this episode, agent $8$ is the center node.}
    \label{fig:traj}
\end{figure*}

\subsection{Localization scheduling}

In the following, we propose three localization scheduling schemes, which depend on the AoI and VoI of each node.

We denote the node scheduled to exploit the $k$-th localization opportunity, occurring at time $kT_s$, as $s(k)$.
We can then formalize the concept of AoI. At any given time $t$, the AoI $\Delta_i(t)$ of node $i$ is the time elapsed since the position of the agent was last determined:
\begin{equation}\label{eq:aoi}
    \Delta_i(t)=\min_{k\in\mathbb{N}:s(k)=i\wedge kT_s\leq t}t-kT_s.
\end{equation}
We then consider the VoI $\theta_i(t)$, which we can first define as its positioning error:
\begin{equation}~\label{eq:voi_naive}
    \theta_i(t)=\mathbb{E}\left[\|p_i-\hat{p}_i\|^2\right].
\end{equation}
Using the characterization of the error in~\eqref{eq:error}, we can rewrite the definition as a function of the AoI $\Delta_i(t)$, using the properties of the $\chi^2$ distribution:
\begin{equation}\label{eq:voi_age}
 \theta_i(\Delta_i)=d\sigma_i^2\Delta_i.
\end{equation}

Finally, we can consider a more complex model of the VoI, which fully exploits the bar-and-joint framework $\mathcal{F}(\mathcal{G})$: we compute the distance-based centrality of agent $i$, $\zeta_i$, as
\begin{equation}
    \zeta_i=\frac{n-1}{\sum_{j\in\mathcal{N}_i}d_{ij}}.
\end{equation}
We can then weight each node's VoI by its centrality, giving a higher value to nodes that are closer to multiple others in the ideal formation:
\begin{equation}~\label{eq:voi_central}
    \rho_i(t)=\zeta_i \mathbb{E}\left[\|p_i-\hat{p}_i\|^2\right].
\end{equation}

\begin{table}[b]
\captionsetup{font=small, width=\textwidth}
\caption{Relevant simulation parameters}.
\centering
\scriptsize{\begin{tabular}{cc|cc|cc}
\toprule
Parameter & Value & Parameter & Value & Parameter & Value\\ \midrule
$N_{\text{ep}}$ & $100$ & $T$ & $10$~s & $T_s$ & $0.1$~s\\
$K_E$ & $100$ &$K_P$ & $10$ & $K_F$& $50$\\
$d,n$ & $3,8$ &$\sigma_0$ & $0.5$~m/s & $d_0$& $5$~m\\
%$\sigma_0$ & $0.5$~m/s &&&&\\
\bottomrule
\end{tabular}\label{tab:param}}
\end{table}

We can then provide the three scheduling algorithms: firstly, \emph{Maximum Age First (MAF)}, which selects the node with the highest AoI. In our scenario, this corresponds to a simple round-robin scheduling, as we assume that the localization process is ideal. The scheduled node is simply
    \begin{equation}
        s(k)=\argmax_{i\in\mathcal{V}}\,\Delta_i(kT_s).
    \end{equation}

We can then define the \emph{Maximum Expected Error (MEE)} scheduler, which considers the naive definition of VoI in~\eqref{eq:voi_naive}. Thanks to the relation in~\eqref{eq:voi_age}, this can also be expressed as a function of the AoI, so as to select node
    \begin{equation}
    s(k)=\argmax_{i\in\mathcal{V}}\theta_i(kT_s)=\argmax_{i\in\mathcal{V}}\,d\sigma_i^2\Delta_i(kT_s).
    \end{equation}
Finally, the \emph{Maximum Value (MV)} scheduler considers the centrality of the node as well, adopting the more advanced VoI definition from~\eqref{eq:voi_central}:
    \begin{equation}
    s(k)=\argmax_{i\in\mathcal{V}}\rho_i(kT_s)=\argmax_{i\in\mathcal{V}}\,d\zeta_i\sigma_i^2\Delta_i(kT_s).
    \end{equation} 
We note that the expected positioning error is independent from the actual position of each node, and so is its centrality, which only depends on the formation. All scheduling schemes can then be pre-computed, and as long as the agents receive periodic synchronization beacons, scheduling information does not need to be broadcast unless the system operator changes the desired formation.

We remark that the communication cost is independent from the scheduler, as the various schemes just allocate the same communication resources to different nodes.

\begin{figure*}[t]
    \centering
    \begin{subfigure}[b]{0.48\textwidth}  % Adjust width as needed
        \centering
        \includegraphics[width=\textwidth]{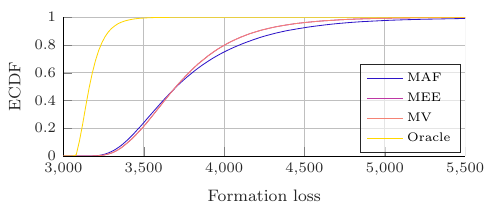}
        \caption{Symmetrical formation.}
        \label{fig:ccdf_sym}
    \end{subfigure}
    \hfill
    \begin{subfigure}[b]{0.48\textwidth}  % Adjust width as needed
        \centering
        \includegraphics[width=\textwidth]{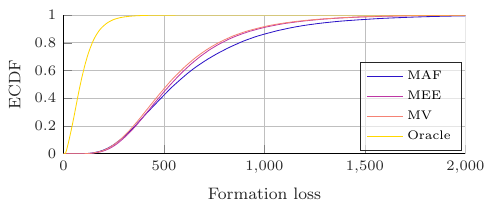}
        \caption{Asymmetrical formation.}
        \label{fig:ccdf_asym}
    \end{subfigure}
    \caption{ECDF of the formation loss over $p$, removing the first second of each episode.}
    \label{fig:ccdf}
\end{figure*}

\section{Numerical results}
\label{sec:exp}

\begin{figure*}[t]
    \centering
    \begin{subfigure}[b]{0.48\textwidth}  % Adjust width as needed
        \centering
        \includegraphics[width=\textwidth]{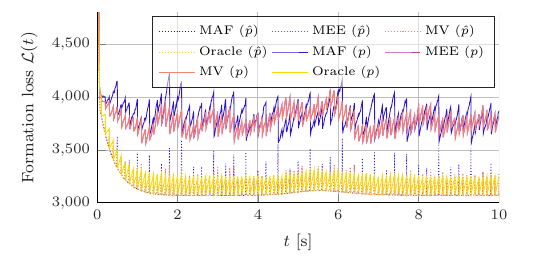}
        \caption{Symmetrical formation.}
        \label{fig:loss_sym}
    \end{subfigure}
    \hfill
    \begin{subfigure}[b]{0.48\textwidth}  % Adjust width as needed
        \centering
        \includegraphics[width=\textwidth]{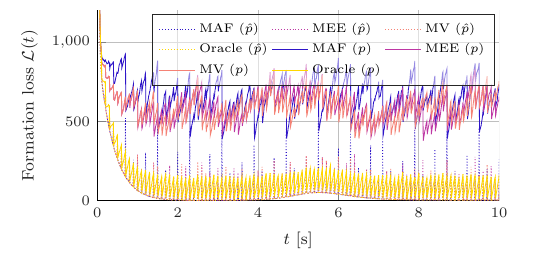}
        \caption{Asymmetrical formation.}
        \label{fig:loss_asym}
    \end{subfigure}
    \caption{Average formation loss over time for the three considered scheduling algorithms, as well as the oracle. Dotted lines indicate the estimated loss (i.e., the one over $\hat{p}_i$), solid lines indicate the actual loss over $p_i$.}
    \label{fig:loss}
\end{figure*}

We evaluated the proposed scheme by performing a Monte Carlo simulation over $N_{\text{ep}}$ episodes lasting $T=10$~s each.\footnote{The simulation code is available at $\quad\quad\quad\quad\quad\quad\quad\quad\quad\quad\quad$ \url{https://github.com/marcofabris92/voi-formation-control}} The full system parameters are available in Table~\ref{tab:param}. 

We consider a set of $n=8$ agents, which aim at respecting two possible formations in three-dimensional space ($d=3$): a symmetric one in which all nodes are arranged on the vertices of a cube whose side is $d_0=5$~m, and an asymmetric one in which one of the nodes is in the barycenter of the cube, so that there is a missing vertex. The distances $d_{ij}$ can be easily computed geometrically. Nodes have the following noise standard deviation:
\begin{equation}
    \sigma_i=\sigma_0(1+i),
\end{equation}
with $\sigma_0=0.5$~m/s. Nodes' positions in the formation are randomly rearranged in each round.

The desired formation centroid trajectory to be tracked is yielded by $p_{c,des}(t) = \begin{bmatrix}
    t, & 0, & 5-5\tanh{(t-T/2)}
\end{bmatrix}^\top$~m; whereas, the components $p_i(0)$ of the initial position $p(0)$ for the ensemble of agents are generated as $p_k(0) = (10 + 0.1 \varsigma_k)$~m, where $\varsigma_k$, $k=1,\ldots,N$, is a zero-mean Gaussian random variable with unit variance. The edge set $\mathcal{E}$ of each formation is reported in Table~\ref{tab:edgesets}, while the node weights and edge weights are assigned respectively as
$a_{ii} := \zeta_i / \sum_{j\in \mathcal{V}} \zeta_j$ and $a_{ij} := \sqrt{a_{ii}a_{jj}}$ for all $(i,j) \in \mathcal{E}$.

Aside from the MAF, MEE, and MV scheduling schemes, we consider an \emph{Oracle} case in which all agents are localized every $T_s$: naturally, this represents an upper bound to system performance, as it involves using $n$ times as much bandwidth as any of our schemes.

The trajectories of the nodes following the asymmetrical formation in a random episode when using different scheduling schemes are shown in Fig.~\ref{fig:traj}: we can easily see that the trajectories of nodes $1$ and $5$ are smoother than those of other nodes, as they have a lower noise.  However, the oracle scheme shown in Fig.~\ref{fig:traj_asym_or} has a lower noise, while the trajectory obtained when using the MAF scheduler, shown in Fig.~\ref{fig:traj_asym_maf}, requires more corrections and strays slightly farther from the ideal one than the one in Fig.~\ref{fig:traj_asym_mv}, which is the result of using the MV scheduler.

We can get a better idea of the performance of the schemes by considering the formation loss $\mathcal{L}(t) = \mathcal{L}(p(t),\mathcal{F}(\mathcal{G})) $, which we define as
\begin{equation}
    \mathcal{L} = \frac{K_P}{2} \|p_{c,\text{des}}-p_c\|^2  
  +\frac{K_F}{2}\!\!\!\sum_{(i,j)\in\mathcal{E}} \!\!\!a_{ij}  \left(\|p_i-p_j\|^2-d_{ij}^2\right)^2.
\end{equation}
%\begin{align}
%  \mathcal{L}(p) =~ & \frac{K_P}{2} \|p_{c,\text{des}}-p_c\|^2  \nonumber\\
%  &+\frac{K_F}{2}\sum_{(i,j)\in\mathcal{E}} a_{ij}  \left(\|p_i-p_j\|^2-d_{ij}^2\right)^2.
%\end{align}
The formation loss over time, averaged over the $N_{\text{ep}}$ episodes, is shown in Fig.~\ref{fig:loss}. As all nodes have the same centrality $\zeta_i$ in the symmetric formation, MEE and MV have an almost identical trend in Fig.~\ref{fig:loss_sym}, while there is a noticeable difference in Fig.~\ref{fig:loss_asym}, representing the loss when targeting the asymmetrical formation. In this case, the loss is more stable when using MV, while MEE tends to oscillate more. However, the peaks and troughs generated by MAF are significantly more pronounced, showing that the performance of age-based scheduling is much more unstable. In all cases, computing the loss on the estimated position $\hat{p}$ leads to significantly lower value than the actual loss, with peaks corresponding to the localization of one of the agents. On the other hand, the oracle scheme maintains an actual loss that is only slightly higher than the loss computed using $\hat{p}$, as each node is able to get much more information about its position and correct its course. We also note that the initial loss  is very high, as nodes start out far from their assigned positions, but rapidly decreases, as seen in the trajectories in Fig.~\ref{fig:traj}.

Finally, Fig.~\ref{fig:ccdf} shows the Empirical Cumulative Distribution Function (ECDF) of the formation loss, excluding the first second of each episode: the plots show that the VoI-aware policies are able to obtain a better performance than MAF in both scenarios. However, MEE and MV are equivalent in the symmetrical formation, as shown in Fig.~\ref{fig:ccdf_sym}, while Fig.~\ref{fig:ccdf_asym} shows that MV has a slight advantage when using the asymmetrical formation, as it weights each node's priority by its centrality.

\section{Conclusions}
\label{sec:conc}
In this work, we presented a joint framework for distributed formation control and network-based localization: our scheme relaxes the assumption that each node is able to know its position with perfect accuracy, and considers a scheme in which it obtains it through triangulation with multiple BSs. We present state-of-the-art AoI- and VoI-based schedulers, showing that a greater awareness of the application can help improve performance, particularly when considering more difficult tasks such as an asymmetrical formation. Future work on the subject may focus on the further optimization of the consensus schemes regulating the exchange of position information, jointly designing control and communication strategies.

%\begin{ack}
%Place acknowledgments here.
%\end{ack}

\bibliography{ifacconf}

%\appendix
%\section{A summary of Latin grammar}
%\section{Some Latin vocabulary}

\end{document}